\documentclass[12pt,epsf]{article}
\usepackage{graphicx,amsmath,amssymb}
\begin{document}
%%%%%%%%%%%%%%%%%%%%%%%%%%%%%%%%%%%%%%%%%%%

\def\a{\alpha}
\def\b{\beta}
\def\c{\varepsilon}
\def\d{\delta}
\def\e{\epsilon}
\def\f{\phi}
\def\g{\gamma}
\def\h{\theta}
\def\k{\kappa}
\def\l{\lambda}
\def\m{\mu}
\def\n{\nu}
\def\p{\psi}
\def\q{\partial}
\def\r{\rho}
\def\s{\sigma}
\def\t{\tau}
\def\u{\upsilon}
\def\v{\varphi}
\def\w{\omega}
\def\x{\xi}
\def\y{\eta}
\def\z{\zeta}
\def\D{\Delta}
\def\G{\Gamma}
\def\H{\Theta}
\def\L{\Lambda}
\def\F{\Phi}
\def\P{\Psi}
\def\S{\Sigma}

\def\o{\over}
\def\beq{\begin{eqnarray}}
\def\eeq{\end{eqnarray}}
\newcommand{\gsim}{ \mathop{}_{\textstyle \sim}^{\textstyle >} }
\newcommand{\lsim}{ \mathop{}_{\textstyle \sim}^{\textstyle <} }
\newcommand{\vev}[1]{ \left\langle {#1} \right\rangle }
\newcommand{\bra}[1]{ \langle {#1} | }
\newcommand{\ket}[1]{ | {#1} \rangle }
\newcommand{\EV}{ {\rm eV} }
\newcommand{\KEV}{ {\rm keV} }
\newcommand{\MEV}{ {\rm MeV} }
\newcommand{\GEV}{ {\rm GeV} }
\newcommand{\TEV}{ {\rm TeV} }
\def\diag{\mathop{\rm diag}\nolimits}
\def\Spin{\mathop{\rm Spin}}
\def\SO{\mathop{\rm SO}}
\def\O{\mathop{\rm O}}
\def\SU{\mathop{\rm SU}}
\def\U{\mathop{\rm U}}
\def\Sp{\mathop{\rm Sp}}
\def\SL{\mathop{\rm SL}}
\def\tr{\mathop{\rm tr}}

\def\IJMP{Int.~J.~Mod.~Phys. }
\def\MPL{Mod.~Phys.~Lett. }
\def\NP{Nucl.~Phys. }
\def\PL{Phys.~Lett. }
\def\PR{Phys.~Rev. }
\def\PRL{Phys.~Rev.~Lett. }
\def\PTP{Prog.~Theor.~Phys. }
\def\ZP{Z.~Phys. }

%%%%%%% added by Fumi %%%%%%%%%%
% FROM HERE
%\newcommand{\beq}{\begin{equation}}   
%\newcommand{\eeq}{\end{equation}}
\newcommand{\bea}{\begin{eqnarray}}   
\newcommand{\eea}{\end{eqnarray}}
\newcommand{\bear}{\begin{array}}  
\newcommand {\eear}{\end{array}}
\newcommand{\bef}{\begin{figure}}  
\newcommand {\eef}{\end{figure}}
\newcommand{\bec}{\begin{center}}  
\newcommand {\eec}{\end{center}}
\newcommand{\non}{\nonumber}  
\newcommand {\eqn}[1]{\beq {#1}\eeq}
\newcommand{\la}{\left\langle}  
\newcommand{\ra}{\right\rangle}
\newcommand{\ds}{\displaystyle}
\def\SEC#1{Sec.~\ref{#1}}
\def\FIG#1{Fig.~\ref{#1}}
\def\EQ#1{Eq.~(\ref{#1})}
\def\EQS#1{Eqs.~(\ref{#1})}
\def\TEV#1{10^{#1}{\rm\,TeV}}
\def\GEV#1{10^{#1}{\rm\,GeV}}
\def\MEV#1{10^{#1}{\rm\,MeV}}
\def\KEV#1{10^{#1}{\rm\,keV}}
\def\lrf#1#2{ \left(\frac{#1}{#2}\right)}
\def\lrfp#1#2#3{ \left(\frac{#1}{#2} \right)^{#3}}
\def\REF#1{Ref.~\cite{#1}}
\newcommand{\osc}{{\rm osc}}
\newcommand{\ed}{{\rm end}}
\newcommand{\iso}{{\rm NL}^{\rm (ISO)}}
\def\dda#1{\frac{\partial}{\partial a_{#1}}}
\def\ddat#1{\frac{\partial^2}{\partial a_{#1}^2}}
\def\dd#1#2{\frac{\partial #1}{\partial #2}}
\def\ddt#1#2{\frac{\partial^2 #1}{\partial #2^2}}
\def\lrp#1#2{\left( #1 \right)^{#2}}
% UNTIL HERE

%%%%%%%%%%%%%%%%%%%%%%%%%%%%%%%%%%%%%%%%%%%%%%%%%%%%%%%%%%%%%%%%%%%%

\baselineskip 0.7cm

\begin{titlepage}

\begin{flushright}
TU-933\\
\end{flushright}

\vskip 1.35cm
\begin{center}
{\large \bf 
Isocurvature Constraints and  Anharmonic Effects  \\
on  QCD Axion Dark Matter
}
\vskip 1.2cm
Takeshi Kobayashi,$^{a,b}$\footnote{takeshi@cita.utoronto.ca},
Ryosuke Kurematsu,$^{c}$\footnote{rkurematsu@tuhep.phys.tohoku.ac.jp}
and Fuminobu Takahashi$^{c}$\footnote{fumi@tuhep.phys.tohoku.ac.jp}

\vskip 0.4cm

{ \it $^a$Canadian Institute for Theoretical Astrophysics,
 University of Toronto, \\ 60 St. George Street, Toronto, Ontario M5S
 3H8, Canada}\\
{\it $^b$Perimeter Institute for Theoretical Physics, \\ 
 31 Caroline Street North, Waterloo, Ontario N2L 2Y5, Canada}\\
{\it $^c$ Department of Physics, Tohoku University, Sendai 980-8578, Japan}

\vskip 1.5cm

\abstract{ 
We revisit the isocurvature density perturbations induced by quantum fluctuations of the axion field
by extending a recently developed analytic method and approximations to a time-dependent scalar potential,
which enables us to follow the evolution of the axion until it starts to oscillate. 
We find that, as the initial misalignment angle approaches the hilltop of the potential, 
the isocurvature perturbations become significantly enhanced, while the non-Gaussianity
parameter increases slowly but surely.   As a result, the isocurvature constraint
on the inflation scale is tightened as $H_{\rm inf} \lesssim \mathcal{O}(100)$\,GeV for the axion decay constant $f_a \lesssim10^{10}$\,GeV, 
near the smaller end of the axion dark matter window. 
We also derive useful formulae for the power spectrum and non-Gaussianity of the isocurvature
perturbations. 
}
\end{center}
\end{titlepage}

\setcounter{page}{2}

\section{Introduction}
The identity of dark matter is one of the central issues in cosmology and particle physics.
The QCD axion is an interesting and plausible candidate for dark matter; it is
a Nambu-Goldstone (NG) particle associated with the spontaneous breakdown of the
Peccei-Quinn (PQ) symmetry  introduced to solve the strong CP problem in QCD~\cite{Peccei:1977hh,QCD-axion}. 
As the axion settles down at the potential minimum, the CP violating
phase $\theta$ is set to a vanishingly small value, solving the strong CP problem.

The dynamical relaxation of the CP phase necessarily induces
coherent oscillations of the axion, which contribute to dark matter, since the axion is stable in a cosmological time scale
for the decay constant  in the axion window~\cite{Raffelt},
\beq
\GEV{9} \;\lesssim f_a \;\lesssim \GEV{12}.
\label{window}
\eeq
The lower bound comes from astrophysical constraints including the cooling argument of globular-cluster stars,
and the upper bound from the requirement that the axion density should not
exceed the observed dark matter density for the initial misalignment angle of order unity.
If the fine-tuning of the initial position is allowed, or if non-standard cosmology is assumed~\cite{Kawasaki:1995vt},
the upper bound can be relaxed to e.g. the GUT or string scale.

One of the features of the axion is that its quantum fluctuation during inflation naturally induces
an almost scale-invariant isocurvature density fluctuation, which would leave a distinctive imprint on the CMB 
spectrum.\footnote{
If the QCD interactions become strong at an intermediate or high energy scale in the very early Universe,
the axion may acquire a sufficiently heavy mass, leading to suppression of
the isocurvature perturbations~\cite{Jeong:2013xta}.
}
The observed CMB spectrum is known to be fitted extremely well by a nearly scale-invariant
adiabatic density perturbation, and  a mixture of the isocurvature
perturbations is tightly constrained by the observations~\cite{Hinshaw:2012fq,Ade:2013rta}. 
This constraint can be interpreted as an upper bound on the inflation energy density as the quantum fluctuation of the axion field
is set by the Hubble parameter during inflation~\cite{Seckel:1985tj,Lyth:1989pb}. 

The statistical information contained in the density fluctuations can be exploited by
estimating higher order correlation functions. The non-Gaussianity of the isocurvature 
density fluctuations has been studied from both theoretical~\cite{Kawasaki:2008sn,Langlois:2008vk,Kawakami:2009iu,
Langlois:2011zz,Langlois:2010fe}  and observational~\cite{Hikage:2008sk,Hikage:2012be,Hikage:2012tf} point of view.
In particular, the non-Gaussianity of the CDM isocurvature perturbation induced by the axion
was estimated under an approximation assuming a quadratic potential for the axion~\cite{Kawasaki:2008sn}. 
This approximation however breaks down as the initial misalignment angle deviates from the
CP symmetric vacuum, and  an anharmonic effect must be taken into account.
The anharmonic effect on the axion abundance and the isocurvature power spectrum was studied in 
Refs.~\cite{Turner:1985si,Lyth:1991ub,Strobl:1994wk,Bae:2008ue,Visinelli:2009zm}. 
However, while there is a consensus on the anharmonic effect on the axion abundance, there are
apparent disagreement as to how the isocurvature perturbations are affected by the anharmonic effect.

Recently  an analytic method and approximations to compute density perturbations induced
by a curvaton field with a broad class 
of the potential was developed by Kawasaki and two of the present authors (Kobayashi and Takahashi) (KKT  in the following)~\cite{Kawasaki:2011pd}.
Using this new method, we can analytically estimate the density perturbations generated in
a curvaton mechanism~\cite{Linde:1996gt,Enqvist:2001zp,Lyth:2001nq,Moroi:2001ct}  for a potential significantly deviated from the
simple quadratic one~\cite{Kawasaki:2011pd,Kobayashi:2012ba,Kawasaki:2012gg}.\footnote{
In the hilltop limit,  the density fluctuations receive a dominant contribution from the fluctuations of the timing when the 
curvaton starts to oscillate, rather than from those of the oscillation amplitude, in contrast to what is usually assumed for the quadratic potential.
To our knowledge, this observation was first made in Ref.~\cite{Kawasaki:2008mc}, and its analytic understanding was developed in Ref.~\cite{Kawasaki:2011pd}. See also Ref.~\cite{Kobayashi:2013bna} for spectator field models in light of the spectral index after the Planck. 
} The analytic and numerical results
were found to agree with each other to a very good accuracy, typically, within several percent.
This method was however limited to a time-independent potential for the curvaton.

%In this paper we revisit the isocurvature perturbations
%induced by the axion field fluctuations, and estimate the anharmonic effect on the isocurvature perturbations
%have not been studied so far. 
%

In this paper we  extend the analytic method to a time-dependent potential, 
and then apply the method to the case of the QCD axion. 
We find it essential to define the onset of oscillations properly in order to evaluate
the axion isocurvature perturbations.
To this end we solve an attractor equation of motion for the axion until it starts to oscillate. 
This is the key to understand the anharmonic effect on the behavior of isocurvature perturbations. 
Based on the extended KKT method, we first correctly estimate the power spectrum and non-Gaussianity 
of the axion isocurvature perturbations for any axion initial position, resolving the aforementioned disagreement 
in the literature. 
%The anharmonic effect on the non-Gaussianity of 
We find that both power spectrum and non-Gaussianity  are enhanced as the initial field
value approaches the hilltop of the potential, thus giving an extremely tight constraint on the inflation scale
for the axion decay constant $f_a = {\cal O}(10^{9-10})$\,GeV,  near the smaller end of the axion dark 
matter window (\ref{window}). We will also provide useful formulae for the power spectrum and
non-Gaussianity of the axion isocurvature perturbations.

Here let us summarize the comparison of our results with the works in the past. First of all,
the axion abundance has been studied extensively in the past, including the hilltop limit,
and we obtained results which are consistent with the previous works. 
Namely, the axion abundance is enhanced as the initial field value approaches the hilltop of the potential.
However, concerning the
power spectrum of the axion isocurvature perturbations, we find that
there are some inconsistencies among the literature. As we shall see later, we find that the power spectrum 
of the axion isocurvature perturbation
gets significantly enhanced toward the hilltop, and its behavior agrees well with the result of 
Ref.~\cite{Lyth:1991ub}  valid near the hilltop.  However, such an enhancement was not
correctly taken into account in Ref.~\cite{Visinelli:2009zm}, which resulted in a much weaker
constraint on the inflation scale for $f_a = {\cal O}(10^{9-10})$\,GeV in their Fig.~1.
In Ref.~\cite{Wantz:2009it}, it was pointed out that the power spectrum can be suppressed or enhanced by
the anharmonic effects depending on the parameters. We do not confirm such suppression in our analysis. 
Furthermore, their enhancement factor is weaker than what we find.

The rest of this paper is organized as follows. In Sec.~\ref{sec:2} we  summarize the basic properties
of the QCD axion. 
In Sec.~\ref{sec:3}, we give a brief review of the recently
developed analytic approach and extend it to include the time-dependent potential. We apply
this analysis to the QCD axion, and calculate the derivative of the e-folding with respect
to the axion field value at the horizon exit. In Sec.~\ref{sec:4}, using the results in Sec.~\ref{sec:3},
we estimate the power spectrum and non-Gaussianity of the
isocurvature perturbations, and derive constraints on the inflation scale as a function of the
axion decay constant. We also provide useful formulae for the power spectrum and non-Gaussianity
of the axion isocurvature perturbations; given the analytic expression for the axion dark matter abundance, 
one can easily estimate them using the formulae.
The last section is devoted to  conclusions.

%%%%%%%%%%%%%%%%%%%%%%%%%%%%%%%%%%%%%%%%%
\section{Axion Dark Matter}
\label{sec:2}
%%%%%%%%%%%%%%%%%%%%%%%%%%%%%%%%%%%%%%%%%
We here briefly review the basics of the QCD axion dark matter. The axion is a NG
 boson associated with the spontaneous breaking of the PQ symmetry, which was introduced
to solve the strong CP problem.  Throughout this paper we assume that the PQ symmetry is 
already broken during inflation and is not restored after inflation\footnote{
 The isocurvature perturbations are not generated if the PQ symmetry is restored during or after inflation~\cite{Linde:1990yj,Lyth:1992tx}.
In this case, however, the topological defect such as the cosmic strings and domain walls
are generated, and in particular, the PQ sector must be such that the domain wall number $N_{DW}$
is unity.
}, and that the PQ breaking scale remains unchanged during and after inflation. 
See Ref.~\cite{Linde:1991km,Kasuya:2009up} for the case where the PQ breaking scale, and therefore  the coefficient of the kinetic term for the axion,
significantly evolves after inflation, which would suppress/enhance the axion quantum fluctuations. A similar effect is possible 
if there is a non-minimal coupling to gravity~\cite{Folkerts:2013tua}. It is also possible to make the axion so heavy during
inflation that its quantum fluctuations are suppressed if the QCD becomes strong at an intermediate or 
high energy scale~\cite{Jeong:2013xta}.

The axion has a flat potential protected by the PQ symmetry, which however is explicitly
broken by the QCD anomaly.  Thus, as the comic temperature drops down to the QCD
scale,  $\Lambda_{\rm QCD} \approx 200$\,MeV, the QCD interactions become strong and 
non-perturbative, and the axion gradually acquires a finite mass from the QCD instanton effects. 
The axion potential is approximately given by
\beq
V(a,T) \;=\; m_a(T)^2 f_a ^2 \left(1-\cos\lrf{a}{f_a}\right) 
\eeq
where $f_a$ is the axion decay constant and $m_a(T)$ is the temperature dependent axion mass
approximately given by~\cite{Preskill:1982cy}
\beq
m_a(T) \;\approx\; 
\left\{
\bear{cc}
\ds{\lambda m_0 \lrfp{\Lambda_{\rm QCD}}{T}{p}}
 &~~{\rm~for~} T \gg \Lambda_{\rm QCD}\\
m_0 & ~~{\rm~for~} T \ll \Lambda_{\rm QCD}\\
\eear
\right.
\eeq
with $\lambda \approx 0.1$ and $p \approx 4$. We set the CP symmetric
vacuum to be the origin of the axion. 
The axion mass at zero temperature is related to the decay constant as
\beq
m_0 \;\approx\; \frac{\sqrt{z}}{1+z} \frac{m_\pi f_\pi}{f_a} \simeq 6.0 \times 10^{-6}  {\rm \, eV} \lrf{\GEV{12}}{f_a},
\eeq
where $z \equiv m_u/m_d$,  $m_\pi =135$\,MeV, $f_\pi = 92$\,MeV,  and we used the conventional value for
$z = 0.56$ in the second equality. 

At a sufficiently high temperature, the axion mass is much smaller than the Hubble parameter.
When the temperature becomes as low as $T \lesssim {\cal O}(1)$\,GeV, the axion starts to 
oscillate around the potential minimum. Throughout this paper we assume the radiation dominated
Universe when the axion starts to oscillate. 
For a small initial misalignment angle, $\theta_* = a_*/f_a \ll \pi/2$, 
the axion potential can be approximated with a quadratic potential. Then the axion abundance is
given by
\beq
\Omega_a h^2 \;\simeq\; 0.2\, \theta_*^2 \lrfp{f_a}{\GEV{12}}{\frac{7}{6}},
%\Omega_a h^2 \;\simeq\; 0.7 \lrfp{\theta_*}{\pi}{2} \lrfp{f_a}{\GEV{12}}{\frac{7}{6}},
%\Omega_a h^2 \;\simeq\; 0.195\, \theta_*^2 \lrfp{f_a}{\GEV{12}}{1.184},
\eeq
where $h$ is the present-day Hubble parameter in units of $100$ km s$^{-1}$ Mpc$^{-1}$.
One can see from this form that the observed dark matter abundance can be naturally 
explained by the coherent oscillations of the axion for the initial misalignment angle
of order unity and for $f_a \simeq \GEV{12}$. On the other hand,
the initial misalignment angle must be finely tuned to be
an extremely small value for $f_a \gg \GEV{12}$ to avoid the overclosure of the Universe.

If the axion field initially sits near the maximum of the potential, i.e., $\theta_* \simeq \pi$, 
the commencement of the coherent oscillations can be significantly delayed. This is known
as the anharmonic effect~\cite{Turner:1985si,Lyth:1991ub}.
In the following we  study its effect on the isocurvature fluctuations 
of the axion by using the recently developed analytic method. We shall see that the anharmonic
effect leads to the enhancement of the abundance as well as the isocurvature perturbations of the axion.

%%%%%%%%%%%%%%%%%%%%%%%%%%%%%%%%%%%%%%%%%
\section{Analytical method of density perturbations}
\label{sec:3}
%%%%%%%%%%%%%%%%%%%%%%%%%%%%%%%%%%%%%%%%%
\subsection{Basic strategy}
Here we provide a basic strategy of the analytical method to compute the density
perturbations generated by quantum fluctuations of a scalar field. See Ref.~\cite{Kawasaki:2011pd}
for details. 

The density perturbations induced by a light scalar field depend on the scalar evolution
during and after inflation.  Suppose that the scalar potential is approximated by a quadratic
potential, and that its mass is sufficiently light during and for some time after inflation.
Then the scalar field hardly evolves and stays more or less at the initial position until 
it starts to oscillate. When the Hubble parameter becomes comparable to the mass,
it starts to oscillate, and importantly,
the timing does not depend on the position. This is no longer the case 
for a potential of the general form, and one needs to follow the evolution of the scalar
field until the commencement of coherent oscillations in order to compute the
density perturbations. To this end, we first note that the scalar evolution can be
actually well described by an attractor equation of motion. Then, by using the attractor equation of 
motion, we express the dependence of the e-folding number on the initial position of the scalar field.
 Finally we  compute the power spectrum and non-Gaussianity parameter of the isocurvature
perturbations, making use of the $\delta N$-formalism
\cite{Starobinsky:1986fxa,Sasaki:1995aw,Wands:2000dp,Lyth:2004gb}. The most important ingredients 
are $\partial N/\partial a_*$ and $\partial^2 N/\partial a_*^2$, where $N$ is the e-folding number between the horizon exit and
some time after the matter-radiation equality, and $a_*$ is the field value at the horizon exit of the CMB
scales. The purpose of the rest of this section is to express $\partial N/\partial a_*$  and $\partial^2 N/\partial a_*^2$
in terms of the scalar potential and the axion field.

In Ref.~\cite{Kawasaki:2011pd}, the scalar potential was assumed to be time-independent. 
Here we shall extend the attractor equation of motion to include a possible time-dependence of the
scalar potential, which is an essential feature of the axion.

\subsection{The attractor equation of motion}
Here we  derive an attractor equation of motion of the following form,
\beq
c H \dot{a} + V'(a,t) \;=\;0,
\label{att}
\eeq
where $c$ is a positive constant, a dot represents a $t$-derivative, and a prime
a partial differentiation with respect to the scalar field $a$.
In the following discussion we identify $a$ with the QCD axion, but most of the results in this section can be
straightforwardly applied to a generic scalar field with a
potential~$V(a,t)$ with explicit time dependence. 

The attractor equation for a time-independent
potential was derived in Ref.~\cite{Chiba:2009sj} and also in
Appendix A of Ref.~\cite{Kawasaki:2011pd}. The point is that, 
the scalar evolution can be well described by a first order differential
equation when the curvature of the potential is much smaller than the
Hubble parameter. The coefficient $c$ is determined so that
the attractor equation is consistent with the true equation of motion, $\ddot{a} + 3H \dot{a} + V' = 0$.

To simplify our analysis we consider a potential of the form, $V(a, t) = f(t) v(a)$,
with $\dot{f}(t) =- x H f(t)$, where $x$ is a constant. 
Then the constant~$c$ in (\ref{att}) should satisfy
\beq
c \;=\; 3 - \frac{\dot{H}}{H^2} -x -
 \frac{f v''}{cH^2}.
\eeq
%%%
%%%%
This shows that in a Universe with constant $\dot{H} / H^2$
(i.e. constant equation of state parameter $w = p / \rho$), then as long as
that the potential curvature is as small as $ |f v''| \ll  c^2 H^2$, the
constant $c$ is given by
\begin{equation}
 c \; \approx \; 3 - \frac{\dot{H}}{H^2} - x.
\label{c7}
\end{equation}
One can further check that the approximation (\ref{att}) with (\ref{c7})
is actually a stable attractor for $c > 0$.

In the case of axion, the temperature dependence is given by\footnote{
We have confirmed that such approximation is indeed valid until the commencement of
oscillations for the parameter region of our interest. 
For example, the axion mass can approach its zero-temperature value~$m_0$
before the axion starts to oscillate if the axion is located extremely
close to the hilltop (beyond the region studied in this paper), 
however in such case the axion density would exceed the observed dark
matter density.} 
\beq
f(t) \propto T^{-2p},
\eeq
for $T \gg \Lambda_{\rm QCD}$. This leads to 
\beq
\frac{\dot{f}}{f} \;=\; -2p \frac{\dot{T}}{T} = 2p H,
\label{eq10}
\eeq
i.e., $x=-2p \approx -8$, where we have assumed radiation-domination,
$H=1/2t$,  in the second equality.\footnote{The temperature dependence
of the relativistic degrees of freedom~$g_*$ is neglected
in the second equality of~(\ref{eq10}). We remark that within the
temperature range 
$200 \, \mathrm{MeV} \lesssim T \lesssim 1 \, \mathrm{GeV}$ where the
axion starts its oscillations, $g_*$ changes slowly enough such that its
time variation gives at most $\sim 5$ \% modification to the value of~$x$.}
(Having a constant $x =- \dot{f}/ Hf$ greatly simplifies the analysis for axions,
as we will soon see.)
The coefficient $c$ can be approximated with
\beq
c \;\approx\; 5+2p,
\eeq
where we used ${\dot H} = -2 H^2$. The attractor equation of motion will  be valid as long as
\beq
\left|\frac{f v''}{H^2}\right| \;\ll\; c^2 \sim 170.
\eeq
Roughly speaking, the attractor equation of motion holds until the
curvature of the potential becomes about $10$ times as large as the Hubble parameter.

\subsection{The onset of oscillations}
The early stage of the scalar evolution can be described well by the attractor equation,
which however breaks down at a certain point, and the scalar field starts to oscillate around
the potential minimum. The purpose of this subsection is to define the timing of the commencement of
oscillations, $t=t_{\rm osc}$, or equivalently, $H=H_{\rm osc}$, and to calculate its dependence on 
the initial position $a_*$. Specifically, we will calculate $\partial \ln H_\osc^2/\partial a_\osc$ and
$\partial a_\osc/\partial a_*$.

Setting the potential minimum around which the scalar oscillates to $a =
0$, the oscillations are considered to start when
\beq
\left|\frac{\dot{a}}{H a} \right|_\osc = \kappa,
\label{onset}
\eea
where $\kappa$ is a constant of order unity. We will set it to be unity in the numerical calculations, but
we have confirmed that our results remain almost intact as long as $\kappa$ is of order unity. 
Combined with the attractor equation, we obtain
\beq
H_\osc^2 = \frac{1}{\kappa} \frac{V'_\osc}{c\, a_\osc},
\eeq
where it is assumed that the potential is an increasing function of $a$ from the origin to the field
values of interest. 
Differentiating both sides with respect to $a_\osc$, we obtain
\beq
\dda{\osc} \ln H_\osc^2 \;=\; - \frac{1}{a_\osc} + \frac{1}{ v'_\osc} \left(
 v''_\osc +v_\osc' \frac{\dot{f}_\osc }{f_\osc} \dd{t_\osc}{a_\osc} \right)
\eeq
where $f_\osc \equiv f(t_\osc)$ and $v_\osc \equiv v(a_\osc)$, and it should be noted that the potential explicitly depends on the time, which also
depends on $a_\osc$. (In other words, $a_\osc$ and $t_\osc$ are related through \EQ{onset}.)
Using
\beq
\dd{t}{a} \;=\; 
-\frac{1}{4H} \dda{} \ln H^2,
\eeq
which holds in the radiation dominated era,
 we obtain
%%%
\beq
\dda{\osc}  \ln H_\osc^2  \;=\; \frac{4}{a_\osc} X(a_\osc),
\label{hosc}
\eeq
with
\beq
X(a_\osc) \;=\; \frac{a_\osc}{4} \left(1- \frac{x}{4} \right)^{-1}
\left( \frac{v''_\osc }{ v'_\osc}- \frac{1}{a_\osc} \right).
\label{funcX}
\eeq
For later use we rewrite \EQ{hosc} as
\beq
\dd{t_\osc}{a_\osc}\;=\; - \frac{X(a_\osc)}{H_\osc a_\osc}.
\label{dtda}
\eeq
If the scalar potential is quadratic, the function $X$ vanishes. On the other hand, in the hilltop limit,
$X$ becomes much larger than unity. Thus, $X$ is considered to represent the effect of the deviation of the scalar
potential from  a quadratic one. 

Next we calculate $\partial a_\osc/\partial a_*$. 
Integrating the attractor equation  over $a = a_* \sim a_\osc$ and $H=H_* \sim H_\osc$, 
we obtain
\bea
\int_{a_*}^{a_\osc} \frac{1}{v'(a)} da &=& \mathrm{const.} - \int^{H_\osc} \frac{f(H)}{c H \dot{H}} dH,
\label{inta}
\eea
where $f(H)$ actually means $f(t(H))$, and terms that are independent
of~$a_*$ are denoted as const.
We differentiate both sides with respect to $a_*$ to obtain
\bea
\dd{a_\osc}{a_*}  
&=& \left(1 - \kappa X(a_\osc) \right) ^{-1} \frac{v'(a_\osc)}{v'(a_*)}.
\label{aoas}
\eea
The main results of this subsection are (\ref{hosc}) and (\ref{aoas}), which will be used later to calculate $\partial N/\partial a_*$. 

\subsection{Axion number density}
Next we estimate the axion number density. Since the axion mass increases after the onset 
of oscillations, it is the number density that determines the relic abundance of the axion;
the number density decreases in proportional to the inverse of the volume,
and its ratio to the entropy density is fixed. The axion energy density at a later time
can be estimated by multiplying the number density with the zero-temperature mass, $m_0$.

Denoting the mass at the origin on the onset of oscillations as $m_\osc$,
the number density is related to the potential and mass as
\beq
n_{a,\osc} \simeq \frac{V_\osc}{m_\osc}.
\label{number}
\eeq
Precisely speaking, the number density is also dependent on the kinetic energy.
However, one can see that the kinetic energy contribution becomes smaller
as one goes to the hilltop region, since the onset of oscillations are delayed.
This can be seen as follows. The kinetic energy can be estimated as
\beq
{\rm K.E.} \;=\; \frac{1}{2} {\dot a}_\osc^2 \simeq \frac{\kappa^2}{2} H_\osc^2 a_\osc^2,
\eeq
where (\ref{onset}) is used. On the other hand, the potential energy is 
\beq
{\rm P.E.}  \;\sim\; m_\osc^2 a_\osc^2.
\eeq
Thus, for the delayed onset of oscillations, $H_\osc < m_\osc$, the kinetic energy is
smaller than the potential energy.

Differentiating the number density with respect to $a_\osc$, one obtains
\bea
\dda{\osc} \ln n_{a,\osc} &\simeq& \dda{\osc} \ln \frac{V_\osc}{m_\osc} \non\\
&=& \frac{v'_\osc}{v_\osc} + \dd{t_\osc}{a_\osc} \left( \frac{\dot{f}_\osc}{f_\osc} 
- \frac{ \dot{m}_\osc }{m_\osc} \right),
\eea
where it should be noted that $m_\osc$ has no explicit dependence on $a_\osc$,
and so, $m'_\osc=0$.
Using $m_\osc \propto f_\osc^{1/2}$ and \EQ{dtda}, one arrives at
\bea
\dda{\osc} \ln n_{a,\osc} &\simeq& \frac{v'_\osc}{v_\osc} + \frac{x}{2} \frac{X(a_\osc)}{  a_\osc}.
\label{nosc}
\eea

\subsection{The e-folding number}
Now we are ready to calculate the derivative of the e-folding number with respect to $a_*$,
which is directly related to the primordial curvature perturbations in the $\delta N$ formalism. 
We are interested in the e-folding number between the horizon exit of the cosmological scales and some time
after the matter-radiation equality, and these times are represented by $t_*$ and $t_\ed$. Specifically
we take the slicings at $t=t_\osc$ and $t=t_\ed$ as the flat slicing and uniform density slicing, respectively.
The e-folding number between $t_*$ and $t_\ed$ is given by
\bea
N &=& \int_{t_*}^{t_\ed} H(t') dt',
\eea
Let us split the integral into two pieces;
\bea
N_\alpha &\equiv& \int_{t_*}^{t_\osc} H(t') dt',\\
N_\beta &\equiv& \int_{t_\osc}^{t_\ed} H(t') dt'.
\eea
It is useful to use the radiation energy density, or the entropy density, instead of the time.
We adopt the entropy density, $s$,  which scales as the inverse of the volume, 
even if the relativistic degrees of freedom changes with time;
\bea
{\dot s} + 3H s &=& 0.
\eea
The e-folding numbers can be rewritten as
\bea
N_\a &=& - \frac{1}{3} \ln s_\osc + {\rm const.},\non\\
N_\beta &=& - \frac{1}{3} \ln \frac{s_\ed}{s_\osc},
\eea
where $s_\osc = s(t_\osc)$ and $s_\ed = s(t_\ed)$, and the constant term in $N_\a$ does not depend on $a_*$.
The radiation energy density $\rho_r$ is related to the entropy density as
\bea
\rho_r &=& \frac{\pi^2 g_*}{30} T^4 = \frac{\pi^2 g_*}{30} \left(\frac{45}{2\pi^2 g_{*s}} s\right)^{\frac{4}{3}} \propto g_* g_{*s}^{-\frac{4}{3}} s^\frac{4}{3}.
\eea
We are interested in the derivative of the e-folding number with respect to $a_*$. The variation of $a_*$, $\delta a_*$,
is set by the Hubble parameter during inflation, $H_*$, and the corresponding fractional change of the temperature, $\delta T_\osc/T_\osc$,
is much smaller than unity for the parameters of our interest. Thus, we can practically neglect the change 
of $g_*(T_\osc)$ and $g_{*s}(T_\osc)$ due to the variation of $a_*$. 
Similar discussions apply to $g_*(T_{\mathrm{end}})$ and
$g_{*s}(T_{\mathrm{end}})$ as well. 
Then we obtain
\bea
\dd{N_\a}{a_*} &\simeq& -\frac{1}{4} \dda{*} \ln \rho_{r, \osc} \simeq  -\frac{1}{4} \dda{*} \ln H_{\osc}^2, \\
\dd{N_\b}{a_*} &\simeq& -\frac{1}{4} \dda{*}  \ln \frac{\rho_{r,\ed}}{\rho_{r,\osc}} \simeq -\frac{1}{4} \dda{*}  \ln \rho_{r,\ed} + 
\frac{1}{4} \dda{*} \ln H_{\osc}^2,
\label{nb}
\eea
where we have assumed that the Universe was radiation-dominated and the axion and the other CDM components were 
negligible at $t=t_\osc$.
The advantage of using the entropy density  is that
it becomes clear that  the above formulae are still valid
 even if the relativistic degrees of freedom changes between $t_\osc$ and
$t_\ed$.

Now we estimate the first term in \EQ{nb}, which turns out to be rather involved. 
 To this end we define
\bea
\label{R}
R &\equiv &\left. \frac{\rho_c}{\rho_r} \right|_\ed\\
r &\equiv &\left. \frac{\rho_a}{\rho_c} \right|_\ed
\label{r}
\eea
where the total DM density is given by $\rho_c = \rho_m + \rho_a$, and $\rho_m$ denotes the CDM component other than the QCD axion.
Using $R$ and $r$, we have (note that $H_{\mathrm{end}}$ on the final
uniform density slicing is a constant that is chosen independently of~$a_*$)
\bea
\dda{*} \ln \rho_{r, \ed} &=& \dda{*} \ln \left(3 H_\ed^2 M_P^2 - \rho_{m,\ed} - \rho_{a,\ed} \right) \non\\
&=& \frac{1}{\rho_{r,\ed}}\left(- \dd{\rho_{m,\ed}}{a_*} - \dd{\rho_{a, \ed}}{a_*}\right)\non\\
&=& -R(1-r)  \dda{*} \ln{\rho_{r,\ed}} - \dd{R(1-r)}{a_*} - Rr \dda{*} \ln \left(m_0 n_{a,\osc} e^{-3N_\b} \right). \non\\
\eea
We suppose that the entropy perturbation between $\rho_m$ and $\rho_r$
is not generated by the fluctuation of $a_*$. (Hence we consider the CDM
components other than the axion are produced while the axion has little
effect on the expansion of the Universe.) 
Thus,
\bea
\label{ad_dm}
\dda{*}\ln \left(\frac{\rho_m}{s}\right)_\ed &=& 0,
\eea
namely,
\bea
\dda{*} \ln  R(1-r) &=&- \frac{1}{4} \dda{*} \ln \rho_{r,\ed}.
\label{R(1-r)}
\eea
So, we obtain
%%%
\bea
\dda{*} \ln \rho_{r, \ed}
&=&- \frac{4Rr}{4+3R(1-r)}\left( \dd{\ln n_{a,\osc}}{a_*} - 3 \dd{N_\b}{a_*}  \right).
\eea
Substituting this into \EQ{nb}, we arrive at
\bea
\dd{N_\b}{a_*} &=&\frac{Rr}{4+3R} \dd{\ln n_{a,\osc}}{a_*}+ 
\frac{4+3R(1-r)}{4(4+3R)} \dda{*} \ln H_{\osc}^2. 
\eea

To summarize,
\bea
\dd{N}{a_*} 
&=& \frac{Rr}{4+3R} \dda{*} \left(\ln n_{a,\osc} - \frac{3}{4} \ln H_{\osc}^2 \right),\non\\
&=&   \frac{Rr}{4+3R}  \left(1 - \kappa X \right) ^{-1} \frac{v'_\osc}{v'_*} 
\left(\frac{v'_\osc}{v_\osc} - \left(3-\frac{x}{2} \right) \frac{X}{  a_\osc}  \right),
\label{dnda}
\eea
where we used \EQ{hosc} and \EQ{nosc}  in the last equality, and $X$ is understood as $X(a_\osc)$.
This result is reasonable, since it vanishes in the limit of $r \rightarrow 0$, i.e., when the
axion has negligible energy density.

In order to estimate the non-Gaussianity of the isocurvature perturbations, we need
to calculate $\partial^2 N/\partial a_*^2$. The following formulae are useful for this purpose;
\bea
\dd{R}{a_*} &=& 4 (1+R) \dd{N}{a_*},\\
\dd{Rr}{a_*} &=& \left(4+R(3+r) \right)\dd{N}{a_*},
\eea
which can be shown by direct calculation.  After long calculation, we obtain the second derivative of $N$ as
\bea
\ddt{N}{a_*} 
&=&   \frac{16+8R(3-r) + 9 (1-r) R^2}{(4+3R) Rr} \left(\dd{N}{a_*} \right)^2  
+ \frac{Rr}{4+3R}  \dda{*} \left( \frac{4+3R}{Rr} \dd{N}{a_*}  \right)
\label{ddN}\\
&=&
\frac{16+8R(3-r) + 9 (1-r) R^2}{(4+3R) Rr} \left(\dd{N}{a_*} \right)^2  \non\\
&&+ \frac{Rr}{4+3R}  \left(1 - \kappa X \right) ^{-1} \frac{v'_\osc}{v'_*} \left[
 \frac{\kappa X'}{\left(1 - \kappa X \right)^{2}} \frac{v'_\osc}{v'_*} \left(\frac{v'_\osc}{v_\osc} - \left(3-\frac{x}{2} \right) \frac{X}{  a_\osc}  \right)
 \right. \non\\
 &&+\left(1 - \kappa X \right)^{-1} \left( \frac{v''_\osc}{v'_*} 
 -\frac{v''_*}{v'_*} \left(1 - \kappa X \right)
 \right)\left(\frac{v'_\osc}{v_\osc} - \left(3-\frac{x}{2} \right) \frac{X}{  a_\osc}  \right) \non\\
 &&\left.
 + \left(1 - \kappa X \right) ^{-1} \frac{v'_\osc}{v'_*} \left(\frac{v''_\osc}{v_\osc}-\frac{v'_\osc{}^2}{v_\osc^2}  - \left(3-\frac{x}{2} \right) \left(\frac{X'}{  a_\osc} 
 -\frac{X}{a_\osc^2} \right) \right)
 \right].\non\\
\label{ddN2}
\eea
Now we are ready to estimate the isocurvature perturbations and its non-Gaussianity.

%%%%%%%%%%%%%%%%%%%%%%%%%%%%%%%%%%%%%%%%%
\section{The axion isocurvature perturbations and its non-Gaussianity}
\label{sec:4}
%%%%%%%%%%%%%%%%%%%%%%%%%%%%%%%%%%%%%%%%%
\subsection{Power spectrum and bi-spectrum}
Let us start by discussing how the two-point and three-point correlation functions
of the isocurvature perturbations can be expressed in the $\delta N$ formalism~\cite{Starobinsky:1986fxa,Sasaki:1995aw,Wands:2000dp,Lyth:2004gb},
following Ref.~\cite{Kawasaki:2008sn}. 

The isocurvature perturbation $S({\vec x})$ is defined as
\bea
    S({\vec x}) &\equiv & 3 \left(\zeta_{c}({\vec x})-\zeta_{r}({\vec
    x})\right),
\label{Sxdef}
\eea
where $\zeta_{c(r)}$ is the curvature perturbation on the slicing where the dark matter (radiation) is spatially homogeneous. 
Using the  $\delta N$ formalism, $\zeta_{c(r)}$ is obtained 
as the fluctuations in the number of e-folds between an initial flat
slicing (which we take as the time of CMB scale horizon exit) and a
final uniform-$\rho_{c(r)}$ slicing, among different patches of the Universe.
Since we are interested in the difference between $\zeta_c$ and $\zeta_r$, we
only need to compute the contribution to~$\zeta_c$ arising from the
axion field fluctuations~$\delta a_*$. 
Thus (\ref{Sxdef}) can be expressed in terms of the derivative of
the e-folding number as~\cite{Kawasaki:2008sn}\footnote{Here we do not
consider fluctuations of CDM components other than the axion, and
further suppose that there are no mixing terms such as $\partial^2 N /
\partial a_* \partial \phi_*$ where $\phi$ is the inflaton field.}
\bea
    S({\vec x}) &= & 3\left(
            \frac{\partial N}{\partial a_{*}}\delta a_{*}({\vec x})+\frac{1}{2}\frac{\partial ^2 N}{\partial a_{*}^2}(\delta a_{*}^{2}({\vec x})-\langle \delta a_{*}^{2}({\vec x})\rangle) + \cdots \right).
\eea
By taking the final uniform-$\rho_c$ slicing for $N$ to be deep in the matter
dominated era, i.e., $R\rightarrow \infty$, the slicing can be identified with a
uniform total density slicing and thus we can 
use the results from the previous section.\footnote{We should remark
that baryons are not taken into account throughout this 
paper. The simplified treatment is sufficient for giving
order-of-magnitude constraints on the axion and inflationary parameters. 
Here let us also note that baryons can be added to
the above analyses by extending the density ratios defined in (\ref{R})
and (\ref{r}) by
\begin{equation}
R \to \tilde{R} \equiv \frac{\rho_{c+b}}{\rho_r},
\qquad
r \to \tilde{r} \equiv \frac{\rho_{a}}{\rho_{c+b}},
\end{equation}
with $\rho_{c+b}$ denoting the total matter (= CDM + baryons) density.
However $S(\vec{x})$ obtained in this way with the choice of a final
uniform-$\rho_{c+b}$ slicing would give isocurvature perturbations
between radiation and total matter.}

One may wonder about the validity of using the $\delta N$
formalism since the CDM and radiation components exchange energy with
each other, as is clearly seen from the explicit temperature dependence
of the axion mass. 
Here we remark that after the interactions are turned off, the
perturbations~$\zeta_c$ can be computed as the $\delta N$ 
between flat and uniform-$\rho_c$ slicings. Moreover since the Universe
expands uniformly between flat slicings, the initial flat slicing can be
chosen arbitrarily, which we take it to be when the CMB scale exits the horizon 
as we can use the usual estimate for field fluctuations $\delta a_*(k) \simeq H_{\rm inf}/\sqrt{2 k^3}$.

The power spectrum and the bispectrum are usually defined in the momentum space as
\begin{align}
      \langle  S(\vec{k_1})S(\vec{k_2})\rangle &\equiv (2\pi)^3 P_{S}(k_1)\delta^{3}(\vec{k_1}+\vec{k_2}), \\
      \langle  S(\vec{k_1})S(\vec{k_2})S(\vec{k_3})\rangle &\equiv (2\pi)^3 B_S(k_1,k_2,k_3)\delta^{3}(\vec{k_1}+\vec{k_2}+\vec{k_3}),
\end{align}
where the isocurvature perturbation in the momentum space is defined by
    \begin{align}
      S(\vec{k}) =\int\!\!d^3x e^{-i {\vec k} \cdot {\vec x}}S({\vec x}).
    \end{align}   
The momentum dependence of the bispectrum can be approximately factored out;
\begin{align}
    B_S(k_{1},k_{2},k_{3})\equiv f_\iso\Bigl[ P_{S}(k_1)P_{S}(k_2) + 2 {\rm \,perms} \Bigr],
\end{align}
where $f_\iso$ is the non-linearity parameter for the isocurvature perturbations.\footnote{
$f_\iso$ is identical to $f_S$ used in Ref.~\cite{Kawasaki:2008sn}.
} 

Similarly we define the power spectrum of the axion field fluctuations;
\bea
    \langle  \delta a_{*}(\vec{k_1})\delta a_{*}(\vec{k_2})\rangle &=& (2\pi)^3P_{\delta a_{*}}(k_1)\delta(\vec{k_1}+\vec{k_2}) 
\eea
where
\beq
P_{\delta a_{*}}(k) \;=\; \frac{H_{*}^2}{2 k^3}.
\eeq

The power spectrum and the non-linearity parameter of the isocurvature
perturbation can be expressed as, to leading order in the
field fluctuations,\footnote{
It is straightforward to take into account of  higher-order contributions,
which can be relevant if $\delta a_*$ is not sufficiently small compared to $a_*$.
This is the case if one considers a large value of $H_{\rm inf}$ and/or a small value
of $r$.  Note however that the expansion of $\delta N$ in terms of $\delta a$ breaks down
for $\delta a_* \sim a_*$.  When the potential can be approximated
by a quadratic one, an order-of-magnitude estimate of $P_S$
is still possible by simply replacing $\theta_*^2$ with $ \theta_*^2 + \sigma_\theta^2$,
where $\sigma_\theta = H_*/(2 \pi f_a)$.
}
\bea
       P_{S}(k) &=& 9\Bigl(\frac{\partial { N}}{\partial a_{*}}\Bigr)^2P_{\delta a_{*}}(k),\non\\
f_\iso &=& \frac{1}{3}\Bigl( \frac{\partial { N}}{\partial a_{*}}\Bigr) ^{-2}\Bigl( \frac{\partial ^2 { N}}{\partial a_{*}^2}\Bigr),
  \label{al:fs}
\eea
where we consider the axion to have Gaussian field fluctuations.
Let us repeat that in the above expressions, we substitute
the results from the previous section deep in the matter-dominated era,
i.e., $R\rightarrow \infty$.

The constraint from the Planck and the WMAP large-scale polarization data reads~\cite{Ade:2013rta}
\bea
\alpha &\lesssim & 0.041 ~~~(95\% {\rm C.L.}),
\label{alpha}
\eea
where $\alpha$ represents the relative amplitude of the power spectrum of the 
isocurvature perturbation to that of the adiabatic one:\footnote{
Note that $\alpha$ defined here corresponds to $\beta_{\rm iso}/(1-\beta_{\rm iso})$ in Ref.~\cite{Ade:2013rta}.
}
\beq
\alpha \;\equiv\; \frac{P_S(k_0)}{P_{\cal \zeta}(k_0)}
\eeq
with $k_0 \equiv 0.05$\,Mpc$^{-1}$. The  power spectrum of the curvature
perturbations is given by~\cite{Ade:2013lta}
\beq
\Delta^2_\zeta(k_0) \;\equiv\; \frac{k^3}{2\pi^2} P_\zeta(k_0) \simeq  2.2 \times 10^{-9}.
\eeq
The bound on the non-linearity parameter $f_\iso$ was
derived using the WMAP 7-yr data~\cite{Hikage:2012be},
\beq
\alpha^2 f_\iso \;=\; 40 \pm 66,
\label{afs}
\eeq
for the scale-invariant isocurvature perturbations, assuming the absence of non-Gaussianity
in the adiabatic perturbations.\footnote{
We expect that the constraint on $f_\iso$ can be improved considerably  by using the Planck data.
}

\subsection{Axionic isocurvature perturbations and its non-Gaussianity}
Now we are ready to obtain the analytical expression for the isocurvature perturbation and its 
non-Gaussianity generated by the axion, by combining the results in the preceding sections. 

\begin{figure}[t!]
\begin{center}
\includegraphics[scale=0.75]{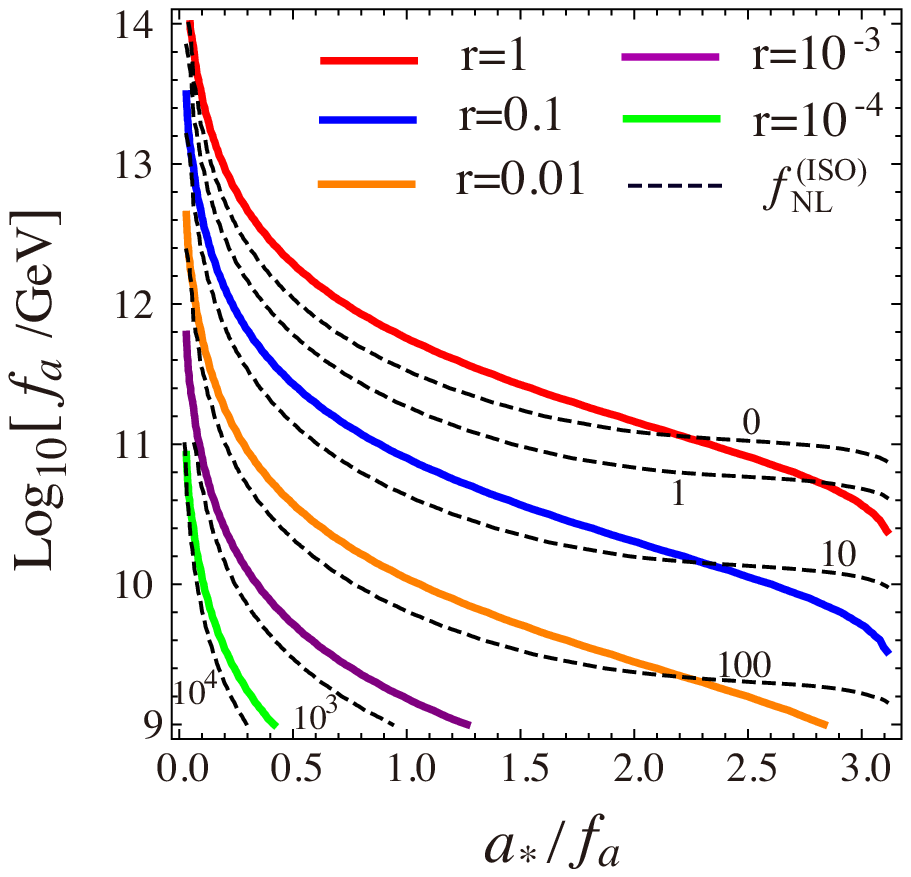}
\includegraphics[scale=0.75]{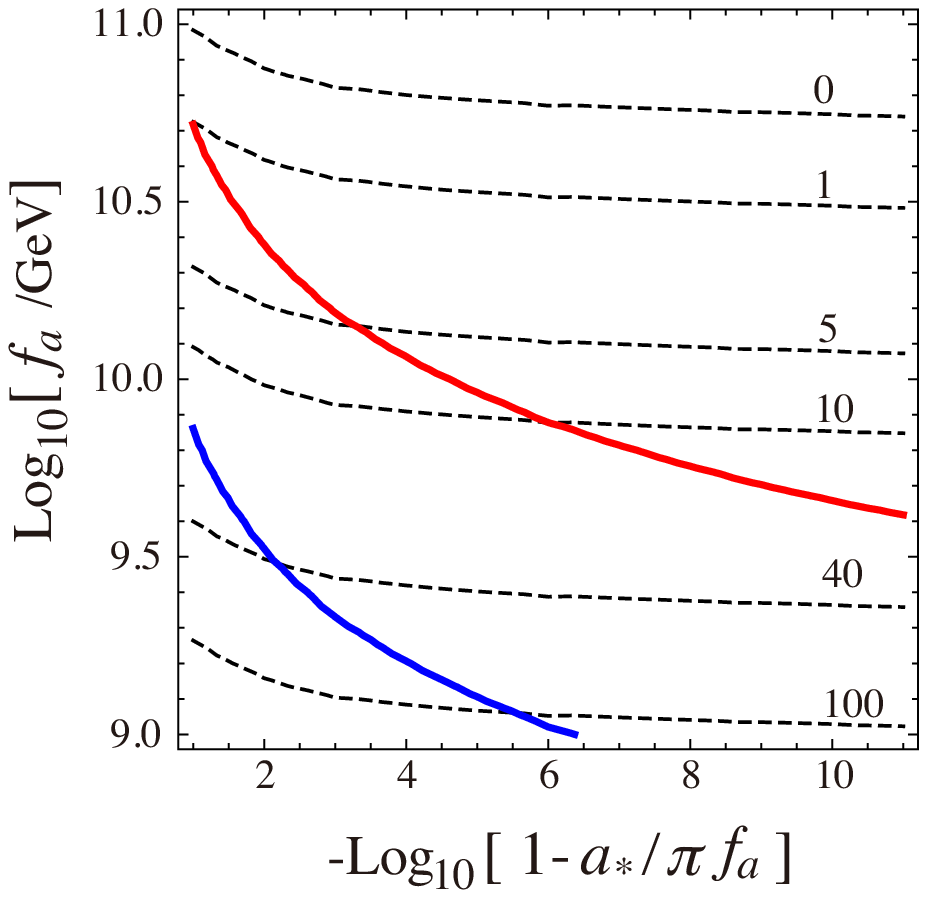}
\caption{
The contours of the fraction of the axion density to the total dark matter density (thick solid lines), $r$, 
and the non-linearity parameter (dashed lines) $f_\iso$.  The contours of $r$ correspond to $r=1, 0.1$, $0.01$, $10^{-3}$ and
$10^{-4}$, from top to bottom. (The contours of $r=0.01, 10^{-3}, 10^{-4}$ are not shown in the right panel.)
}
\label{Oa}
\end{center}
\end{figure}

Let us first calculate the power spectrum of the isocurvature perturbation. 
To simplify the expression, we define $\Delta_S^2(k) \equiv  \frac{k^3}{2 \pi^2} P_S(k)$.
Using \EQ{dnda},  we then obtain
\bea
\Delta_S(k)  &=& 
\left[
\frac{r}{1 - \kappa X} \frac{v'_\osc}{v'_*} 
\left(\frac{v'_\osc}{v_\osc} - \left(3-\frac{x}{2} \right) \frac{X}{
a_\osc}  \right) \frac{H_{*}}{2\pi}
\right]^2
,
\label{ds}
\eea
where we have taken the limit of $R\rightarrow \infty$.
Here, recall that $\kappa$ is a constant of order unity, 
$x = -2p \approx -8$, and $v$ is given as
\begin{equation}
 v(a) \propto 1 - \cos \left( \frac{a}{f_a} \right).
\end{equation}
In the hilltop limit, $X$ (cf. (\ref{funcX})) becomes much larger than unity, and both
$v'_\osc$ and $v'_*$ approach $0$. 
Actually, however,  $v'_\osc$  decreases much more slowly than $v'_*$~\cite{Kawasaki:2011pd}.
Thus, the power spectrum of the isocurvature perturbation  gets  enhanced significantly
by the factor $v'_\osc/v'_*$ , in the hilltop limit. 
Note that $r$ is also enhanced in the hilltop limit (see Fig.~\ref{Oa}), which is much milder compared to the enhancement
due to the factor $v'_\osc/v'_*$.

The non-Gaussianity parameter $f_\iso$ is given by
\bea
f_\iso 
 &=& \frac{1-r}{r}+ \frac{1}{r}  \left(\frac{v'_\osc}{v_\osc} - \left(3-\frac{x}{2} \right) \frac{X}{  a_\osc}  \right)^{-1} \left[ \frac{\kappa X'}{1-\kappa X} + \frac{v''_{\osc}}{v'_{\osc}}
- \frac{v_*''}{v'_\osc} (1-\kappa X)  \right.\non\\
&&
+  \left(\frac{v'_\osc}{v_\osc} - \left(3-\frac{x}{2} \right) \frac{X}{  a_\osc}  \right)^{-1} 
\left(\frac{v''_\osc}{v_\osc}-\frac{v'_\osc{}^2}{v_\osc^2}  - \left(3-\frac{x}{2} \right) \left(\frac{X'}{  a_\osc} 
 -\frac{X}{a_\osc^2} \right) \right].
 \label{fnliso}
\eea
In contrast to the power spectrum, there is no huge enhancement due to $1/v'_*$ in the hilltop limit;
the increase of $f_\iso$ is much milder and is mainly due to $1/v'_\osc$. Note also that $f_\iso$ does not
depend on the inflation scale, while the power spectrum $\Delta_S$ does. 

In order to see the consistency of the above result with the previously known results, let us consider
a quadratic potential, for which $X = 0$ and $X'=0$. This corresponds to the case where the initial
misalignment angle is small. Then one can easily check
\bea
f_\iso & \rightarrow & \frac{1}{2r}-1,
\label{limit}
\eea
in the limit of the quadratic potential, $v(a) \propto a^2$.
This coincides with the result, $f_\iso \simeq 1/2r$ for $r\ll1$ in Ref.~\cite{Kawasaki:2008sn}.

Using the above relations (\ref{ds}) and (\ref{fnliso}), let us estimate the isocurvature perturbations
and its non-Gaussianity of the axion dark matter.  To this end, we need to solve \EQ{onset} (or \EQ{inta}) in order
to evaluate the commencement of oscillations, which determines $a_\osc$, $v_\osc$ as well as $X(a_\osc)$.
We have numerically solved the equation of motion for the axion in a
radiation dominated Universe and obtained $a_\osc$ from \EQ{onset}, then substituted $a_\osc$ to the above relations
to obtain $\Delta_S$ and $f_\iso$.  
The ratio $n_a / s$ is also numerically computed in order to obtain~$r$. 
We present these semi-analytic results in the following.  

In Fig.~\ref{Oa}, we show the contours
of the fraction of the axion to the total dark matter density, $r \equiv \rho_a/(\rho_a+\rho_m)$ (see \EQ{r}), 
and the non-linearity parameter $f_\iso$, in the plane of the initial misalignment angle $a_*/f_a$
and the axion decay constant $f_a$. In the region above the top (red) solid line, the axion abundance
exceeds the observed dark matter abundance.
We can see that, for a given  value of $f_a$,  the axion abundance increases as
the initial position approaches the hilltop, $a_*/f_a = \pi$. Accordingly, the observed dark matter
abundance is realized at a smaller value of $f_a \lesssim \GEV{10}$ for the hilltop initial condition, $1-a_*/\pi f_a \lesssim 10^{-5}$.
The contours of $f_\iso$ show that they are parallel to those of $r$ for $a_*/f \lesssim 1$. This is because, as we have seen above,
$f_\iso$ is determined simply by $r$ when the scalar potential can be approximated with the quadratic potential.
On the other hand, as the initial position approaches the hilltop, the value of $f_\iso$ mildly increases along the contours of $r$.
This mild increase is due to the delayed onset of oscillations, which is one of the features of the hilltop curvaton~\cite{Kawasaki:2011pd}.

\begin{figure}[h!]
\begin{center}
\includegraphics[scale=0.75]{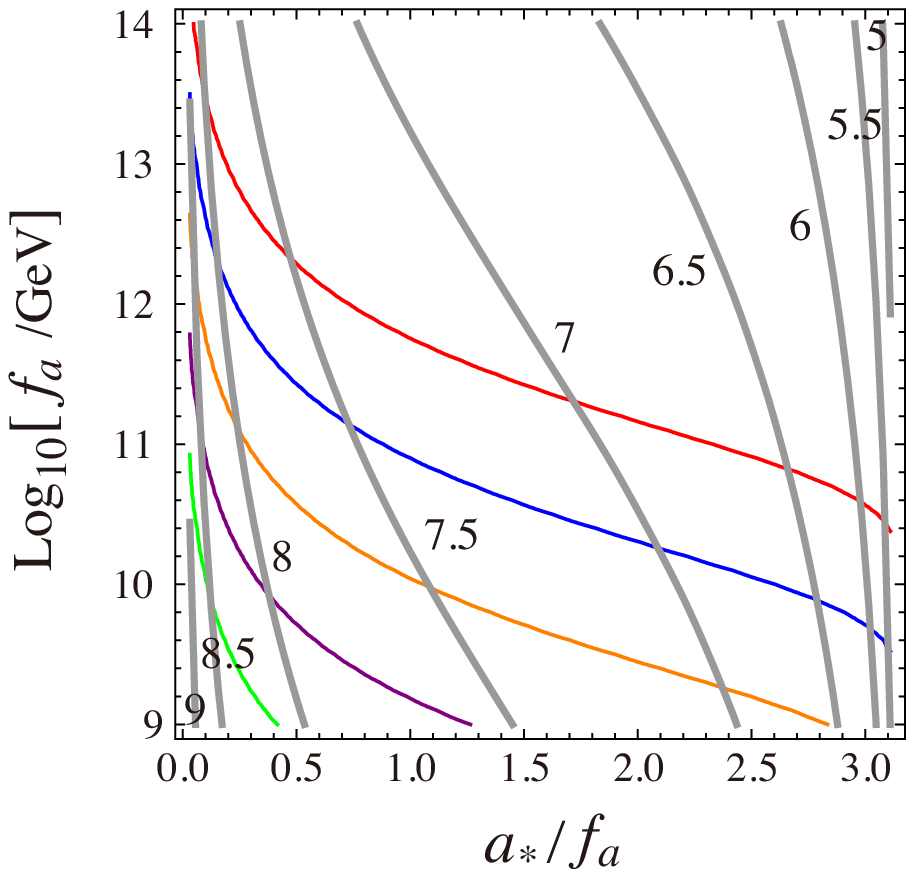}
\includegraphics[scale=0.78]{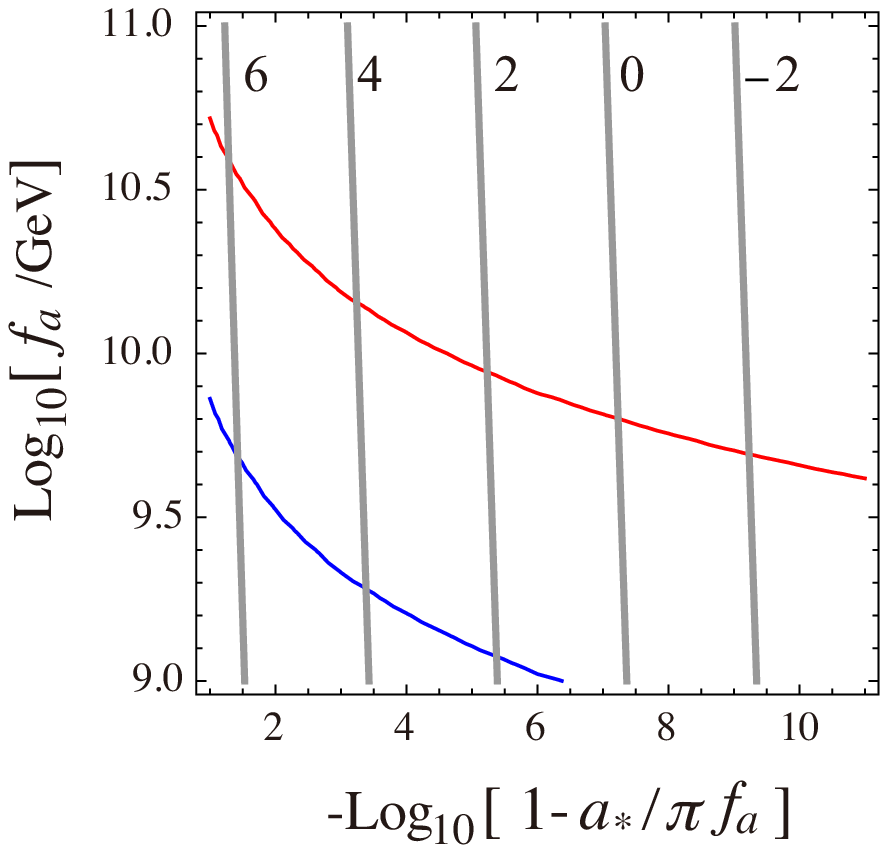}
\caption{
The contours  of the upper bound on the inflation scale, ${\rm Log}_{10}(H_{\rm inf}/{\rm GeV})$, from the
isocurvature perturbation constraint (\ref{alpha}) (thick solid lines). The contours of $r$ in Fig.~\ref{Oa} are shown together.
}
\label{Hinf}
\end{center}
\end{figure}

The inflationary scale, $H_{\rm inf}$, is bounded from above by the isocurvature constraint (\ref{alpha}).
The contours of the upper bound on $\log_{10} (H_{\rm inf}/{\rm \,GeV})$
are shown in \FIG{Hinf}. Since the isocurvature perturbation is significantly enhanced
towards the hilltop, the constraint on $H_{\rm inf}$ becomes tight.  For instance, for $1-a_*/\pi f_a \lesssim 10^{-8}$, 
$H_{\rm inf}$ must be smaller than $\sim 1$\,GeV.\footnote{We have checked that the axion does not fall into
wrong vacuum by climbing over the hilltop of the potential due to the quantum fluctuations, as long as the isocurvature 
constraint on $H_{\rm inf}$ is satisfied.}
On the other hand, the non-Gaussianity constraint (\ref{afs}) gives much weaker constraints in the hilltop region.
One can see this by noting that $f_\iso$ is at most $\simeq 100$ in the
hilltop when $r \gtrsim 0.1$, leading to $\alpha^2 f_\iso \lesssim {\cal O}(0.1)$.
If the axion is a tiny fraction of dark matter (say, $r < 10^{-4}$), 
the non-Gaussianity constraint becomes important, as shown in Ref.~\cite{Kawasaki:2008sn}. 

\begin{figure}[t!]
\begin{center}
\includegraphics[scale=0.75]{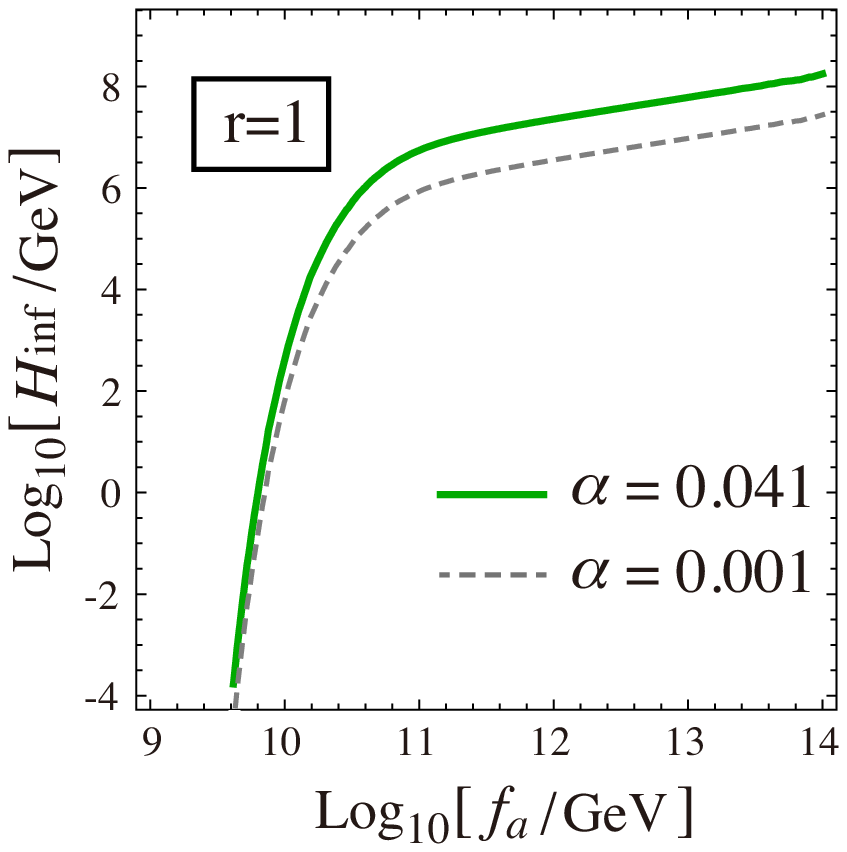}
\includegraphics[scale=0.75]{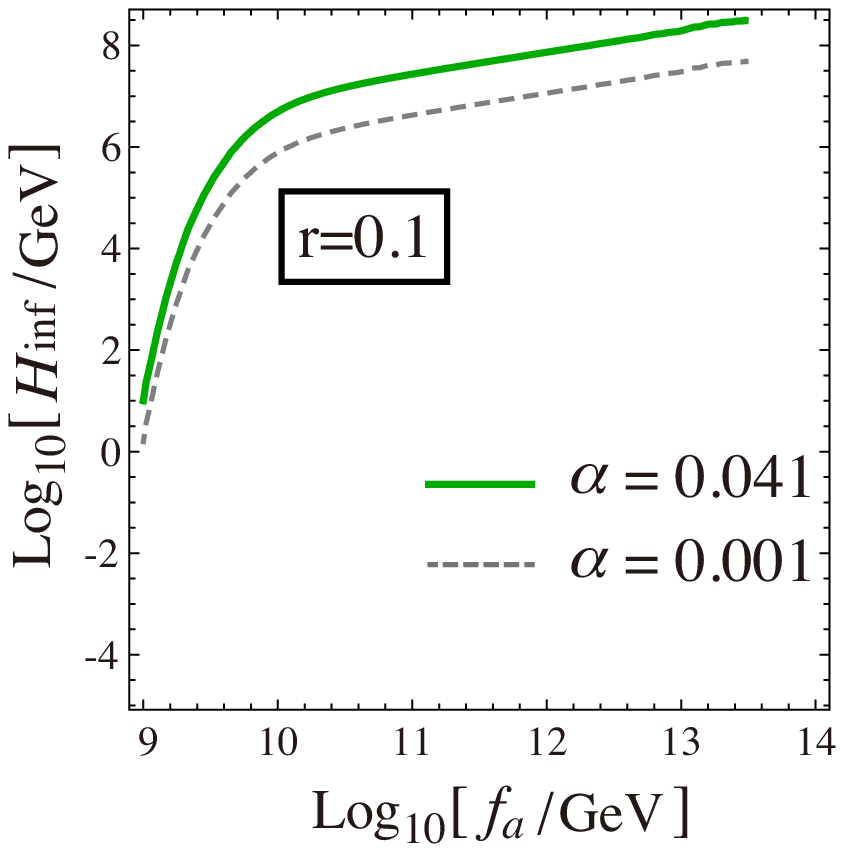}
\includegraphics[scale=0.75]{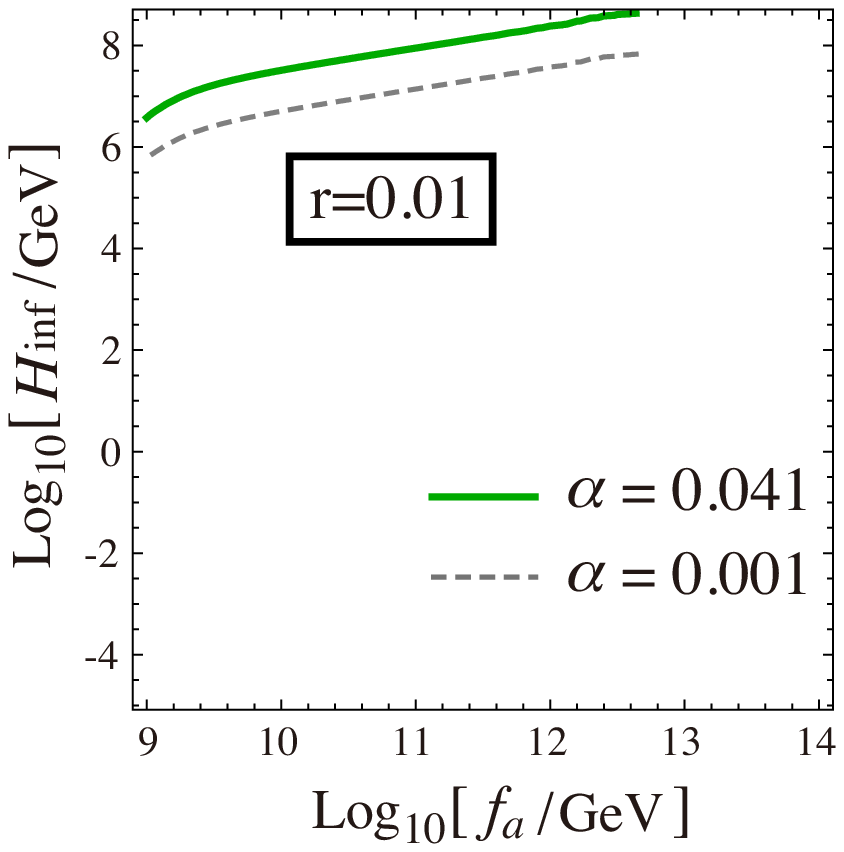}
\caption{
The upper bound on $H_{\rm inf}$ (solid (green) lines) as a function of the axion decay constant $f_a$,
for different fractions of the axion to the total dark matter, $r=1, 0.1,$ and $0.01$.
We used the constraint from the Planck + WMAP large-scale polarization data.
The dashed line is for showing the dependence of the constraints on $\alpha$. 
}
\label{Hinfcon}
\end{center}
\end{figure}

Lastly, we show the constraint on $H_{\rm inf}$ as a function of the decay constant $f_a$ in
Fig.~\ref{Hinfcon}. For comparison, we show two constraints, $\alpha \lesssim 0.041$, and $0.001$.
The first one represents the current constraints from the Planck and WMAP large-scale polarization data, 
while the latter one is just for showing how the constraint changes as the constraint on $\alpha$ improves.
We can see that the constraint becomes extremely tight for $f_a \lesssim \GEV{10}$
due to the anharmonic effect near the hilltop. The effect becomes milder as $r$ decreases since the initial misalignment angles deviates
from the hilltop.

%%%%%%%%%%%%%%%%%%%%%%%%
%%%%%%%%%%%%%%%%%%%%%%%%

%%%%%%%%%%%%%%%%%%%%%%%%
%%%%%%%%%%%%%%%%%%%%%%%%

%%%%%%%%%%%%%%%%%%%%%%%%
%%%%%%%%%%%%%%%%%%%%%%%%

\subsection{Useful expressions for $\Delta_S$ and $f_\iso$}
In order to see if the above semi-analytic estimate correctly describes the power spectrum and non-Gaussianity of the
axion isocurvature perturbations, 
let us compare them with another (more direct) calculation. To this end, it is important to note that (i) the actual abundance
of the QCD axion is much smaller than that of radiation in the early times, which makes it extremely difficult to numerically
solve the evolution of the QCD axion from its formation until the Universe becomes matter-dominated; (ii) in contrast to
the ordinary curvaton scenario, the total dark matter abundance is fixed by observations. Namely, 
the abundance of the dark matter other than the QCD axion must be adjusted so that the total dark matter
 density satisfies $\Omega_c h^2 \simeq 0.12$. The condition for the adiabatic density perturbation, 
 \EQ{ad_dm}, fixes how to compensate the contribution of the QCD axion. 

Let us recap how the isocurvature density perturbations are generated. The axion acquires quantum fluctuations
of order $H_{\rm inf}/2\pi$, which results in the fluctuations of the axion density. The fluctuations of the e-folding
number is through the isocurvature fluctuations of the dark matter density. Therefore, in principle, if we know
how the axion density fluctuates and how the e-folding number fluctuates through the isocurvature fluctuations 
of the dark matter density, we should be able to estimate the isocurvature perturbation and its non-Gaussianity.

The present homogeneous or spatially
averaged dark matter density should satisfy,
\beq
\Omega_c = \Omega_m + \Omega_a \sim 0.27, \label{eq63}
\eeq
and it is $\Omega_a$ that acquires spatial fluctuations, while $\Omega_m$
is independent of $\delta a_*$.
Note however that $\Omega_m$ does depend on $a_*$ through (\ref{eq63}),
but its fluctuations do not  on $\delta a_*$.  More precisely speaking, 
the dependence of $\Omega_m$ on $a_*$ arises from the requirement that the (spatially averaged) total dark matter density
reproduce the observed value. However, its fluctuations is adiabatic and therefore independent of $\delta a_*$. 
Thus, when we expand the e-folding number with respect to $\delta a_*$, we should use
\beq
\dd{\rho_c}{a_*} = \dd{\rho_a}{a_*},
\label{omegaca}
\eeq
where $\rho_c$ and $\rho_a$ are energy densities on a flat slicing at an
arbitrary time. Let us repeat that we are expanding w.r.t. the
fluctuations $\delta a_*$, and for the expansions we need not consider the
spatially averaged observational constraint (\ref{eq63}).

Thus, we may evaluate $\delta N$ as follows;
\beq
\delta N \;=\; \dd{N}{a_*} \delta a_* + \frac{1}{2} \ddt{N}{a_*} (\delta
a_*^2  -\langle \delta a_{*}^{2}\rangle) + \cdots
\eeq
with
\bea
\dd{N}{a_*} &=& \dd{N}{\rho_c} \dd{\rho_a}{a_*},\non\\
\ddt{N}{a_*} &=& \ddt{N}{\rho_c} \left(\dd{\rho_a}{a_*} \right)^2 + \dd{N}{\rho_c} \ddt{\rho_a}{a_*}.
\eea
where we have used \EQ{omegaca}. Thus, the power spectrum and the non-Gaussianity are given by
\bea
 \Delta_S &=& 3  \dd{N}{\rho_c} \dd{\rho_a}{a_*} \frac{H_{\rm inf}}{2\pi},\\
 f_\iso &=& \frac{1}{3} \ddt{N}{\rho_c} \lrp{\dd{N}{\rho_c}}{-2} + 
 \frac{1}{3} \lrp{\dd{N}{\rho_c}}{-1} \lrp{\ddt{\rho_a}{a_*}}{} \lrp{\dd{\rho_a}{a_*}}{-2}.
\eea

The exact solution for the evolution of the Universe
which contains radiation and non-relativistic matter is 
parametrized in terms of a parameter $\xi$ as
\bea
\frac{a_{sf}}{a_{sf0}}&=& \Omega_{c0} \xi^2 + 2 \sqrt{\Omega_{r0}} \xi,\\
H_0 t &=& \frac{2}{3} \Omega_{c0} \xi^3+2\sqrt{\Omega_{r0}} \xi^2,
\eea
where $a_{sf}$ is the scale factor (it should not be confused with the axion $a$),
$t$ is a cosmological time, $\Omega_r$ denotes the density parameter for
radiation, and the subscript ``$0$'' denotes values at an arbitrary time.
In the matter-domination limit, 
\beq
\frac{a_{sf}(t)}{a_{sf0}} &\simeq & \lrfp{3}{2}{\frac{2}{3}} \Omega_{c0}^\frac{1}{3} (H_0 t)^\frac{2}{3}. 
\eeq
The uniform density slicing corresponds to a constant $t$ surface. Thus,
taking both the initial flat slicing and final uniform density slicing to be
in the matter-dominated era, the e-folding~$N$ between them depends
on~$\rho_c$ on the initial slicing as\footnote{The results in this page
can also be derived from the following equation in the matter-dominated era:
\begin{equation}
\rho_a (\vec{x}, t_i) + \rho_m (t_i)  = e^{3 N(\vec{x})} \rho_c (t_f),
\end{equation}
where the energy densities in the left hand side are those on the
initial flat slicing, $\rho_c (t_f)$ is on the final uniform density
slicing, and $N$ is the number of e-folds between the two slicings.}
\bea
\dd{N}{\rho_c} &\simeq& \frac{1}{3 \rho_c}, \\
\ddt{N}{\rho_c} &\simeq&-\frac{1}{3 \rho_c^2},
\eea
giving
\bea
\label{eq74}
\Delta_S &=& \frac{1}{\rho_c} \dd{\rho_a}{a_*} \frac{H_*}{2\pi},\\
 f_\iso &=&-1+ 
\rho_c \lrp{\ddt{\rho_a}{a_*}}{} \lrp{\dd{\rho_a}{a_*}}{-2}.
\label{eq75}
\eea
Since both $\rho_a $ and $\rho_c$ redshift as $\propto a^{-3}$, they can 
be replaced by energy densities on a flat slicing in the later Universe, 
and we finally obtain useful expressions for $\Delta_S$ and $f_\iso$:
\bea
\label{num0}
\Delta_S &=& r \dd{\ln \Omega_a}{\theta_*} \frac{H_*}{2\pi f_a},\\
 f_\iso &=&-1+\frac{1}{r} \lrp{1+ \lrp{\ddt{\ln \Omega_a}{\theta_*}}{} \lrp{\dd{\ln \Omega_a}{\theta_*}}{-2}}{}.
\label{num}
\eea
Thus, given the axion abundance as a function of $a_*$, we can estimate the power spectrum $\Delta_S$
as well as the non-Gaussianity parameter $f_\iso$. (To be
precise, the abundance~$\Omega_a$ in (\ref{num0}) and (\ref{num}) is not
the averaged quantity over the entire observed sky. Instead, it should be
considered as a function of~$a_*$ denoting how the present axion density
responds to the initial axion field value. The critical density here is
taken as a constant.\footnote{Furthermore, the density ratio $r =
\rho_a / \rho_c $ in 
(\ref{num0}) and (\ref{num}) is that on a flat slicing, in
contrast to $r$ in (\ref{r}) defined on a uniform density
slicing. However the two $r$'s are approximately the same and hence
we do not distinguish them in the expressions for 
$\Delta_S$ and $f_\iso$.}) 
If $\Omega_a \propto a_*^2$, these expressions reproduce the known results,
$\Delta_S = r H_*/\pi a_*$, and $f_\iso = -1+1/(2r)$ (see  (\ref{limit})).
Note that our analytic result for $f_\iso$ given in (\ref{fnliso}) is expressed in terms of
the parameters at $t=t_\osc$, while the above estimate (\ref{num}) depends on the final
axion abundance. The advantage of the analytic results in the previous subsection is
that one can easily understand the behavior of the isocurvature perturbation and its non-Gaussianity
in terms of the axion dynamics and the potential.
On the other hand, the expression (\ref{num}) is more useful when the (approximate) analytic
expression for the axion abundance is known. It is also suitable  for numerical calculations.

\begin{figure}[t!]
\begin{center}
\includegraphics[scale=0.7]{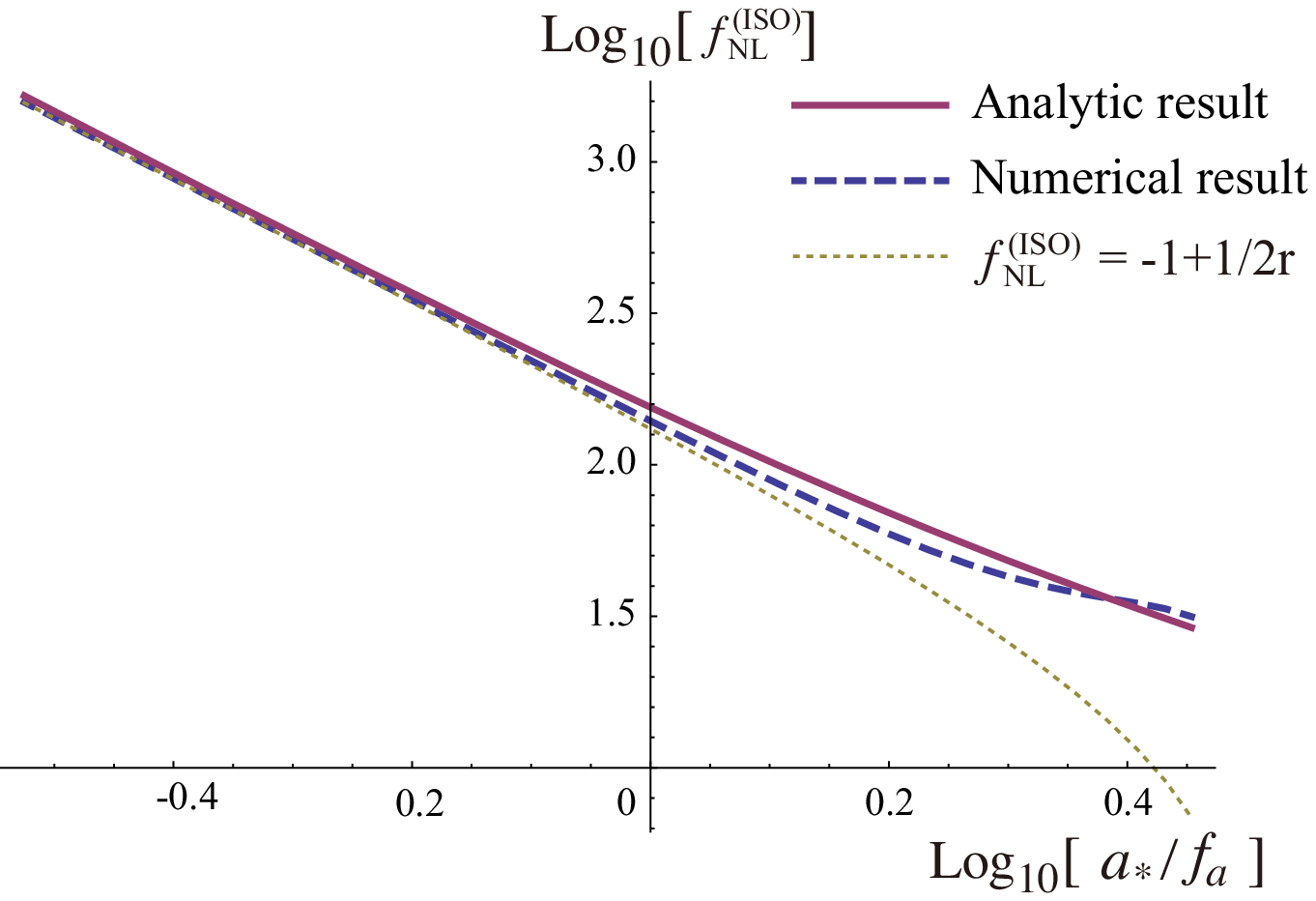}
\includegraphics[scale=0.7]{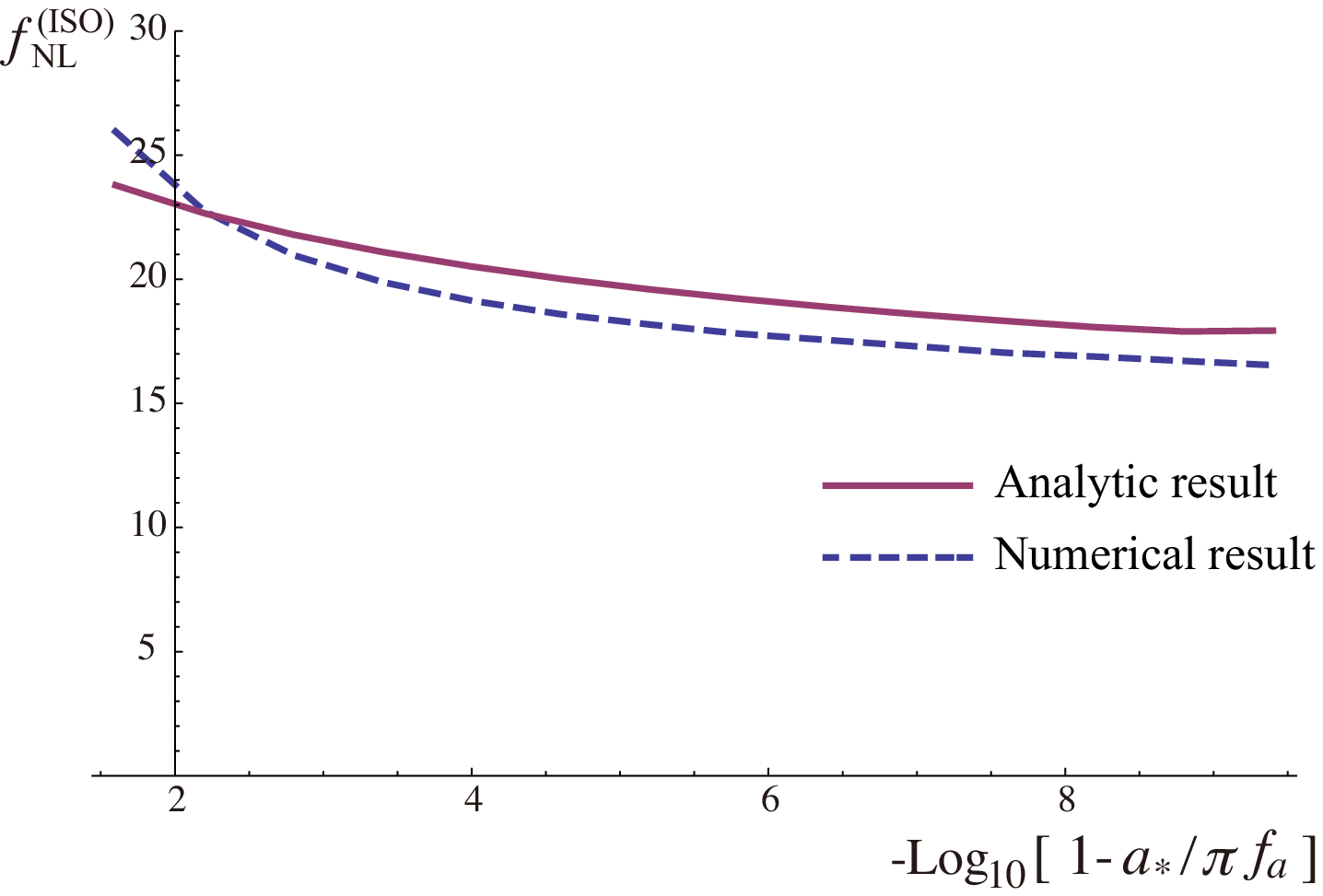}
\caption{The solid (red) lines show the semi-analytic result, while the  dashed (blue) lines represent the numerical results
obtained by using (\ref{num}). We set $f_a = 6\times 10^9$\,GeV. Both agree well with each other
for a wide range of parameters within about 10\%. For comparison, 
the dotted line (yellow) show $f_\iso = -1+1/(2r)$, valid when the potential is approximated with a quadratic potential.
}
\label{f6e9-log}
\end{center}
\end{figure}

In Fig.~\ref{f6e9-log}, we compare two expressions for $f_\iso$,  (\ref{fnliso}) and (\ref{num}),
for $f_a = 6 \times 10^{9}$\,GeV.  They agree with each other well for a wide range of the
initial condition within about $10$\%. We can also see that the expression $f_\iso = -1+1/(2r)$
is indeed valid only near the origin.

%%%%%%%%%%%%%%%%%%%%%%%%
%%%%%%%%%%%%%%%%%%%%%%%%

%%%%%%%%%%%%%%%%%%%%%%%
\section{ Conclusions}
\label{sec:5}
%%%%%%%%%%%%%%%%%%%%%%%

In this paper we have extended the analytic method developed by KKT to  a time-dependent 
potential and then applied it to the QCD axion. We have derived the analytical expressions for
the isocurvature power spectrum (\ref{ds}) and its non-Gaussianity parameter $f_\iso$ (\ref{fnliso}).
Interestingly, the power spectrum significantly increases in the hilltop limit, while the non-Gaussianity 
increases only mildly. Specifically, $f_\iso$ becomes a few tens in the hilltop limit when the axion
is the dominant component of dark matter. If the axion is a subdominant component of dark matter,
it increases as $1/2r - 1$ in the non-hilltop region (see (\ref{limit})). The upper bound on the inflation scale $H_{\rm inf}$
becomes extremely tight in the hilltop limit; $H_{\rm inf} \lesssim  \mathcal{O}(100)\,$GeV for $f_a = \GEV{10}$
when the axion is the dominant component of dark matter. (See \FIG{Hinfcon}).
Our analytic expressions for the axionic isocurvature perturbations will be useful for a probe
of the QCD axion dark matter by using the CMB data.

%%%%%%%%%%%%%%%%%%%%%%%%%%%%%%%%%%%%%
\section*{Acknowledgment}
%%%%%%%%%%%%%%%%%%%%%%%%%%%%%%%%%%%%%
TK thanks Ramandeep Gill and Toyokazu Sekiguchi for helpful conversations. 
 This work was supported by the Grant-in-Aid for Scientific Research on Innovative
Areas (No.24111702, No. 21111006, and No.23104008), Scientific Research (A)
(No. 22244030 and No.21244033), and JSPS Grant-in-Aid for Young Scientists (B) 
(No. 24740135) [FT]. This work was also supported by World Premier International Center Initiative (WPI
Program), MEXT, Japan

\appendix

\section{Another derivation of upper bounds on $H_{\rm inf}$ and evaluation of $f_\iso$}
\begin{figure}[htbp]
\begin{center}
\includegraphics[scale=0.75]{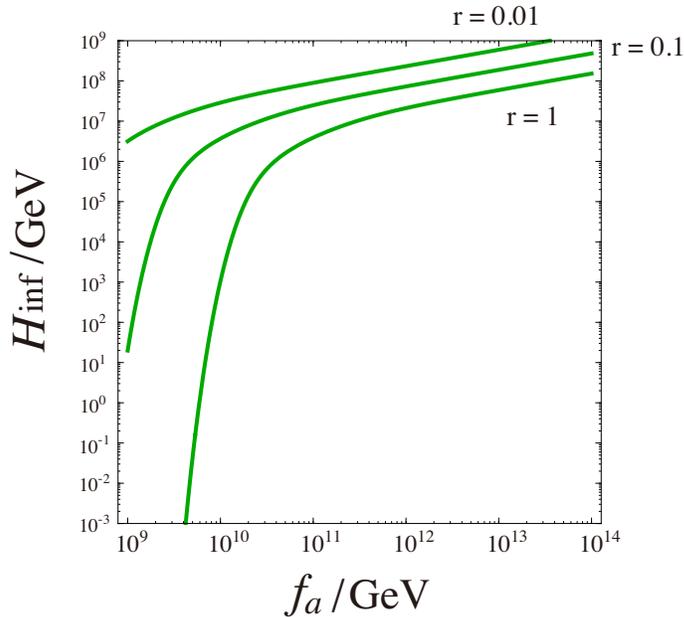}
\caption{
The upper bound on $H_{\rm inf}$ (solid (green) lines) as a function of the axion decay constant $f_a$,
for different fractions of the axion to the total dark matter, $r=1, 0.1,$ and $0.01$, obtained by
using (\ref{num0}) and (\ref{omegaann}).
We used the constraint from the Planck + WMAP large-scale polarization data.
}
\label{Hinfann}
\end{center}
\end{figure}

\begin{figure}[htbp]
\begin{center}
\includegraphics[scale=0.75]{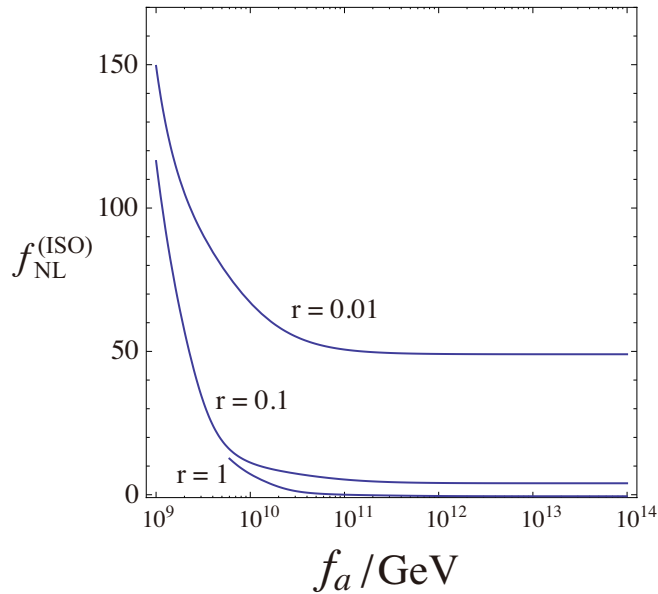}
\caption{
$f_\iso$ for $r=1, 0.1,$ and $0.01$, obtained by
using (\ref{num}) and (\ref{omegaann}).
}
\label{fnlann}
\end{center}
\end{figure}

Here let us estimate the isocurvature constraint on $H_{\rm inf}$ by using the analytic expression
(\ref{num0}) and a known analytic approximation of the anharmonic effect. To this end, we use
the axion abundance~\cite{Bae:2008ue},
\beq
\label{omegaann}
\Omega_a h^2 \;\simeq\; 0.195\, \theta_*^2 F(\theta_*) \lrfp{f_a}{\GEV{12}}{1.184},
\eeq
with the anharmonic coefficient~\cite{Visinelli:2009zm}
\beq
F(\theta_*)\;=\; \left[\ln\lrf{e}{1-\frac{\theta_*^2}{\pi^2}}\right]^{1.184},
\eeq
where we have changed the exponent so as to be consistent with the axion abundance,
although this change does not affect our results.
Substituting these into (\ref{num0}), we obtain the resultant isocurvature perturbation
which is constrained by (\ref{alpha}). We can express the constraint as an upper bound on 
$H_{\rm inf}$ as a function of $f_a$ for fixed $r$ (i.e., $\theta_*$ is given as a function of
$f_a$). The result is shown in Fig.~\ref{Hinfann}, where the upper bounds are shown for
$r=1, 0.1$, and $0.01$ from bottom to top. We can see that the constraints are consistent
with what we obtained by using the semi-analytic formulae (see Fig.~\ref{Hinfcon}).
Similarly, we can evaluate $f_\iso$ by substituting the above expressions into (\ref{num}).
See Fig.~\ref{fnlann}. The enhancement toward the lower value of $f_a$ is due to the
anharmonic effect represented by $X$ in (\ref{fnliso}).

%%%%%%%%%%%%%%%%%%%%%%%%
%%%%%%%%%%%%%%%%%%%%%%%%
%%%%%%%%%%%%%%%%%%%%%%%%
%%%%%%%%%%%%%%%%%%%%%%%%

\end{document}